\documentclass[aps,prb,10pt,showpacs,twocolumn,amsmath,amssymb,superscriptaddress,floatfix]{revtex4-1}

\usepackage{graphicx}
\usepackage{amssymb, dsfont}
\usepackage{multirow}
\usepackage{subfigure}		
\usepackage{epstopdf}
\usepackage{graphicx}
\usepackage{epsfig}
\usepackage{times}
\usepackage[breaklinks]{hyperref}

\usepackage[usenames,dvipsnames]{color}			
\definecolor{purple}{rgb}{0.5,0,0.5}

\begin{document}

\title{Chiral Luttinger liquids and a generalized Luttinger's theorem in fractional quantum Hall edges via finite-entanglement scaling}
\author{D\'{a}niel Varjas}
\author{Michael P. Zaletel}
\affiliation{Department of Physics, University of California, Berkeley, California 94720, USA}
\author{Joel E. Moore}
\affiliation{Department of Physics, University of California, Berkeley, California 94720, USA}
\affiliation{Materials Sciences Division, Lawrence Berkeley National Laboratory, Berkeley, CA 94720, USA}
\begin{abstract}
We use bosonic field theories and the infinite system density matrix renormalization group (iDMRG) method to study infinite strips of fractional quantum Hall (FQH) states starting from microscopic Hamiltonians.
Finite-entanglement scaling allows us to accurately measure chiral central charge, edge mode exponents and momenta without finite-size errors.
We analyze states in the first and second level of the standard hierarchy and compare our results to predictions of the chiral Luttinger liquid ($\chi$LL) theory.
The results confirm the universality of scaling exponents in chiral edges and demonstrate that renormalization is subject to universal relations in the non-chiral case. We prove a generalized Luttinger's theorem involving all singularities in the momentum-resolved density, which naturally arises when mapping Landau levels on a cylinder to a fermion chain and deepens our understanding of non-Fermi liquids in 1D.
\end{abstract} 

\maketitle

\section{Introduction}
	The incompressible liquids of two-dimensional electrons that underlie the fractional quantum Hall effect (FQHE) are believed to support excitations with fractional charge and anyonic statistics.
For the ``abelian'' states, including the Laughlin states at fractional filling $\nu = 1/m$ of the lowest Landau level~\cite{laughlin}, the quasiparticles pick up an exchange phase factor that is neither bosonic nor fermionic, as allowed in two-dimensional systems.\cite{leinaasmyrheim}
Experiments on these quasiparticles typically involve edge~\cite{chang,depicciotto,glattli,heiblum,woowon} or dot~\cite{goldman} geometries.
While it is clear that the gapless excitations at an FQHE edge are different from those of a Fermi liquid, it has been challenging to obtain quantitative agreement between microscopic models and predictions of the effective theory.

	We aim to answer two long-standing questions about fractional quantum Hall edge physics by combining recent analytical and numerical advances in mapping the 2D Landau level of an infinitely long cylinder to an unusual 1D fermion chain.
First, we address whether electron correlation functions along an unreconstructed\cite{ChamonWen1994} edge have the universal behavior predicted by the chiral Luttinger liquid ($\chi LL$) theory.~\cite{wen,wenrev1}
If universal, the edge correlation functions are an experimentally accessible probe of the topological order characteristic of the FQHE phase.
We find unambiguous evidence that for maximally chiral edges with excitations moving in only one direction, the equal time electron correlation functions show universal exponents resulting from the bulk topological order, while non-chiral edges have exponents depending on the intra-edge interactions.
In both cases, the subleading edge exponents obey the relations obtained from $\chi LL$ theory.
\begin{figure}[t]
\includegraphics[width=0.5\textwidth]{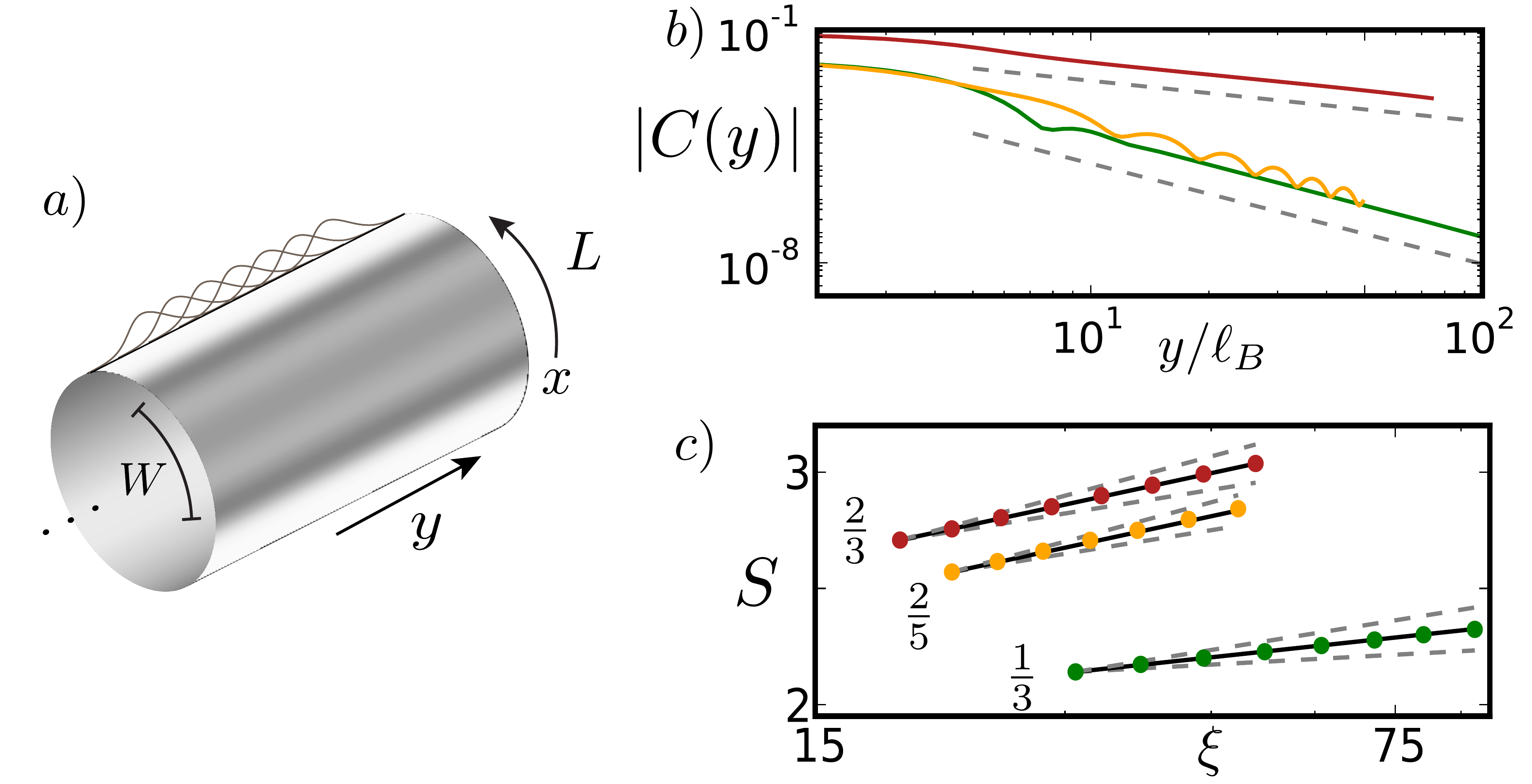}
\caption{\textbf{a) Geometry:} An infinitely long cylinder of circumference $L$ with a trapping potential $V(x)$ squeezing the fluid into a strip of width $W$. Coordinate $x$ runs around the circumference, and $y$ along the infinite length. The Landau orbitals are localized in $y$.
\,\textbf{b) Real space edge correlation functions} $|C(y)| = \langle \psi^\dagger(x, y) \psi(x, 0) \rangle$ for $\nu=1/3$ (green), $\nu = 2/5$ (yellow) and $\nu = 2/3$ (red). $x$ lies near the edge of the strip.
Guiding dashed lines indicate $\eta=3$ and $\eta=1.14$ power laws. 
In the $2/5$ case, oscillations are present due to two $\eta=3$ contributions with different momenta.
\,\textbf{c) Central charge $c$ from finite entanglement scaling.} The correlation length $\xi$ and the entanglement entropy $S$ are measured for increasingly accurate MPS, and are found to scale as $S = \tfrac{c}{6} \log(\xi) + s_0$. The markers are the measured data points; the undashed lines show the scaling relation for $c = \{1, 2, 2\}$ for the $\nu = \{1/3, 2/5, 2/3\}$ states respectively; the dashed lines indicate slopes for $c \pm 1/4$.}
\label{fig:geometry} 
\end{figure}	

	Second, since our method involves mapping the Landau level of a cylinder to a 1D fermion chain, we address how the critical states that arise when studying FQHE edge physics fit into standard descriptions of 1D metals.
For a 1D metal, Luttinger's theorem \cite{Luttinger1960} can be taken to mean that the volume of the Fermi sea, as determined by the non-analytic points in the electron occupation $n_k$, is not modified by interactions (though we note that there is continuing debate over the validity of Luttinger's original formulation in the presence of non-perturbative effects\cite{phillipskane}).
Haldane conjectured ~\cite{haldanenotes, Haldane2009} a `Luttinger sum rule' that extends Luttinger's theorem to FQH strips in the Abelian hierarchy, which was motivated by a simplified picture of the density profile of the hierarchy states.
We unambiguously state and prove a `generalized Luttinger's theorem' which constrains the momenta of singularities in the Greens function of any Abelian FQH state.
In the K-matrix description of the $\chi$LL it takes the simple form $\mathbf{k}^T \mathbf{t} = \pi \nu_T$, where $\nu_T$ is the filling fraction of the cylinder, which agrees with Haldane's conjecture for the hierarchy states.
We prove the constraint using the Lieb-Shultz-Mattis theorem,~\cite{lsm,oshikawaaffleck} and confirm it numerically for some one-component and two-component edges. 
We then clarify why any strip of an $M$th level hierarchy state is, in the 1D picture, an $M$-component Luttinger liquid, which implies that any two states at level $M$ can be adiabatically transformed into each other.

	We use a geometry (Fig.~\ref{fig:geometry}a) in which the edges are infinitely long, which we study using the infinite system density matrix renormalization group (iDMRG).\cite{White-1992, McCulloch-2008} 
Since the problem is translation-invariant along the infinite direction, the approximation in matrix product state numerics is not finite-size but rather finite matrix dimension or ``finite entanglement.''\cite{pollmann2009}
An advantage of this geometry is that correlation functions in the long direction can be obtained over much greater lengths than possible using exact diagonalization, leading to unambiguous scaling behavior.
Recent work on how entanglement scales at conformally invariant quantum critical points~\cite{holzhey,vidal,calabresecardy, pollmann2009} also allows us to extract the central charge of the edge (Fig.~\ref{fig:geometry}c).
The finite circumference does not cut off the correlation length of the edge, and we find that the edge exponents become well quantized at circumferences where there is still significant non-uniformity of the electron density of the bulk.

Tunneling experiments, which probe the frequency and temperature dependence of the edge Green's function, have measured a tunneling exponent of $\alpha \sim 2.7$ for the $1/3$-edge, while the unreconstructed $\chi LL$ prediction is $\alpha = 3$. \cite{Chang1996}
Numerical studies have since investigated the equal time correlation functions and made comparison with $\chi LL$ theory. \cite{PalaciosMacDonald1996,MandalJain2001, Zulicke2003, WanRezayiYang2003,  JoladChangJain2007,HuChenHua2008, JoladJain2009, JoladSenJain2010, Sreejith2011, BernevigBondersonRegnault2012}
There is general agreement that the `model' wave functions obtained from a conformal field theory (CFT), such as the Moore-Read and Laughlin wave functions, must have the exponents predicted by the associated CFT.\cite{PalaciosMacDonald1996, Read2009} 
The situation is much less clear both for the hierarchy states and for longer range interactions.
For the $1/3$-edge there is some evidence, using  finite sized disks, that $\eta_3$ in fact depends on the interactions (Ref.~\onlinecite{MandalJain2001} found $\eta_3 \sim 2.5$ for the Coulomb interaction $r^{-1}$ and $\eta_3 \sim 2.6$ for a Yukawa interaction $r^{-1} e^{-r/\ell_B}$).

\section{Model and Methodology}
\begin{figure}[t]
\includegraphics[width=0.49\textwidth]{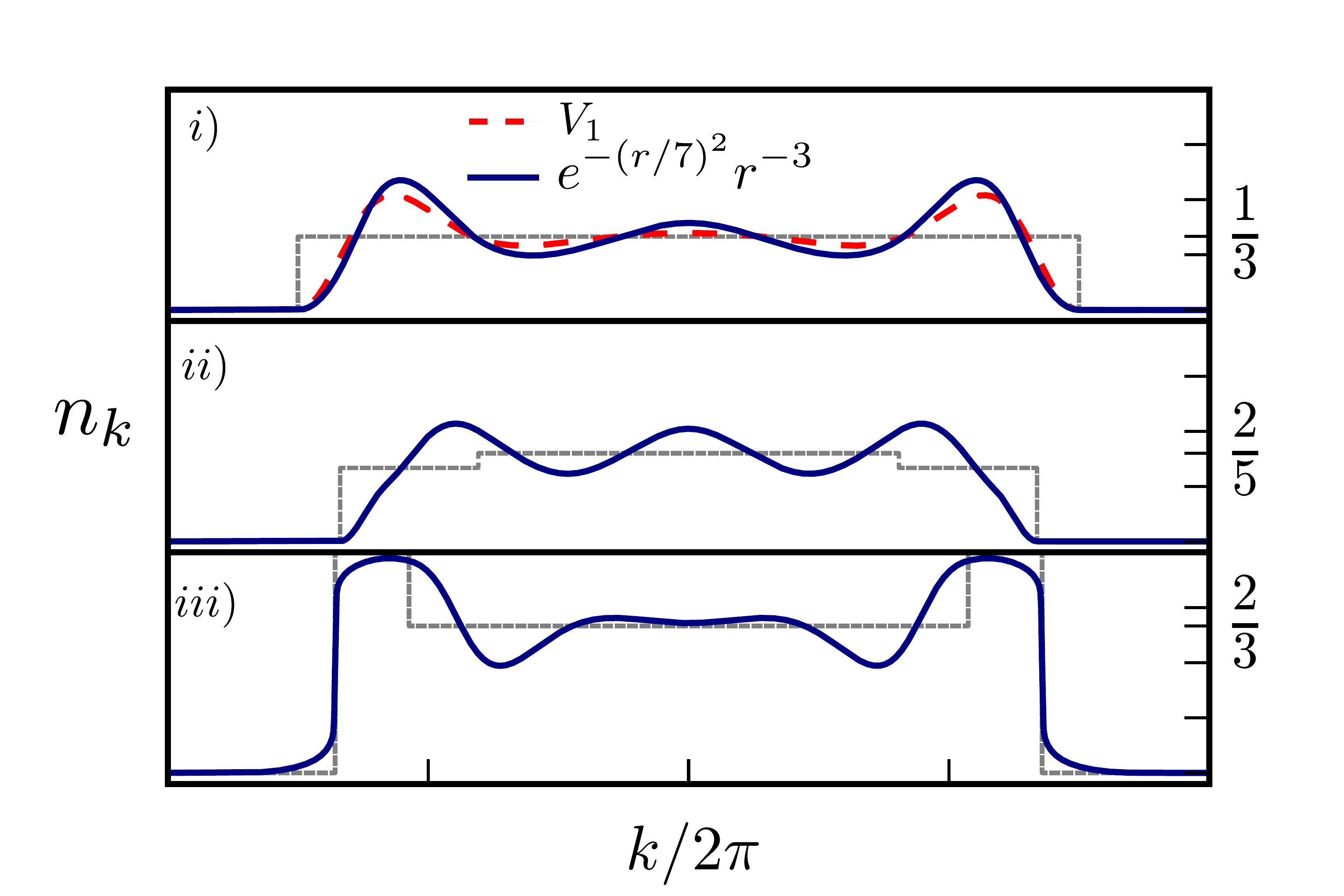}
\caption{
The occupation number $n_k$, which is the Fourier transformation of the orbital correlation function $\langle \psi_n^\dagger \psi_0 \rangle$, or equivalently, the occupation of Landau orbitals for the gauge in which they are localized in $x$.
The gray dashed lines indicate the `wedding cake' caricature of the hierarchy states.
Three states are shown:
\textbf{i}) $\nu = 1/3$ for $W=\tfrac{3}{4} 17 \ell_B$, both for the model `hollow-core' interaction (red dashed) and for a truncated version of the $r^{-3}$ dipole interaction (solid).
\textbf{ii}) $\nu = 2/5$ at $W=\tfrac{5}{8} 19.2 \ell_B$. \textbf{iii}) $\nu = 2/3$ at $W=\tfrac{3}{4} 20 \ell_B$ }.
\label{fig:nk} 
\end{figure}	
	We first map the continuum problem to a translationally invariant fermion chain.\cite{HaldaneRezayi94, BergholtzKarlhede2005, SeidelFu2005}
We start with an infinitely long cylinder of circumference $L$, (Fig.~\ref{fig:geometry}a). Letting $x$ be the coordinate around the cylinder, and $y$ along it, we choose the Landau gauge $\mathbf{A} = B (-y, 0)$ so that each LLL orbital $\phi_n(x,y)$ is localized in the vicinity of 
$y_n = n a$ with $a = 2 \pi \ell_B^2/ L$, 
and can be ordered sequentially.
Interactions and potentials which are translation invariant along $y$ result in a translation invariant Hamiltonian for the fermion chain. 
The interaction matrix elements are nonzero up to infinite distances, but are exponentially suppressed over a characteristic length $\mathcal{O}(L/\ell_B)$ sites. 
The mapping is exact, but we must truncate the interaction terms in order to use the iDMRG, which limits the economically accessible circumference to $L\lesssim 24 \ell_B$.\cite{zaletelmongpollmann}

	After fixing a filling fraction $\nu_T$ for the cylinder, a trapping potential $V(x)$ is projected into the LLL, resulting in hopping terms in the 1D picture.
The trap squeezes the Hall fluid into a denser strip of width $W$.
For example, to obtain the $\nu = 1/3$ strip, we set the overall filling of the cylinder to $\nu_T = 1/4$ and use a trap of width $W = \frac{3}{4}L$.
Rather than realistically modeling a cleaved edge, \cite{WanRezayiYang2003} the trap potentials were chosen to stabilize the desired phases and avoid edge reconstruction, by using a box of depth $t$ and width $W$ convoluted with a gaussian of width $d \sim \ell_B$ to smooth the edge.
For the $1/3$-state, we use a truncated dipolar interaction $r^{-3} e^{-(r/7 \ell_B)^2}$.
For the $2/5, 2/3$-states, we use only the hollow-core Haldane pseudo-potential interaction $V_1$.\cite{HaldaneRezayi94}
		
\subsection{iDMRG and Finite Entanglement Scaling}
	To find the ground state we use the iDMRG method as adapted to the QH problem.\cite{Naokazu-2001, BergholtzKarlhede-2003, zaletelmongpollmann}
The iDMRG algorithm is a variational procedure within the space of infinite matrix product states (iMPS).
An iMPS has finite bipartite entanglement, while the entanglement of the critical edges diverges logarithmically, so the iMPS ansatz cuts off the correlations at a length $\xi_{MPS}$ depending on the dimension $\chi$ of the matrices used.
The finite size effects have been removed, but `finite entanglement' effects introduced.
Analogous to finite size scaling, the finite entanglement ansatz introduces only one length scale, and a `finite entanglement scaling' (FES) procedure has been developed for extracting critical properties. \cite{Tagliacozzo2008, pollmann2009, Pirvu2012, Kjall-2013}
One advantage of FES is that the complexity to obtain a correlation length $\xi_{MPS}$ scales as $\mathcal{O}(\xi_{MPS}^{3 \kappa})$ for an exponent $\kappa$ that depends on the central charge.\cite{pollmann2009}
In contrast, to find the ground state of a disk of circumference $L$ in exact diagonalization (ED) scales as $\mathcal{O}(\alpha^{L^2})$.
We obtain states with $\xi_{MPS} \sim 200 \ell_B$, an order of magnitude larger than the largest disk circumferences obtained from ED, and somewhat larger than the disk circumference obtained through composite fermion ED.
For the largest simulations used here, $\chi \sim 1400$, which required about 30 cpu-hours.

	The output of the iDMRG calculation is an iMPS representation of the  approximate ground state, from which we can efficiently measure  any desired observable.
		
\subsection{Obtaining the edge-exponents}
	The electronic edge exponents are encoded in power-law contributions to the equal time electron correlation function $C(x; y) =  e^{i x y \ell_B^{-2}} \langle \psi^\dagger(x, y) \psi(x, 0) \rangle$. Here $\psi(x, y)$ is the electron operator in the FQH model, and the phase factor is chosen for convenience. 

	We first review the expected form of $C$ in $\chi LL$ theory. The low energy effective theory of a generic abelian FQH edge is described using the $K$-matrix formalism:\cite{wenrev1}
\begin{align}
S = \frac{1}{4\pi}\int dy dt \left(K_{IJ} \partial_t \phi_I \partial_y \phi_J - V_{IJ} \partial_y \phi_I \partial_y \phi_J\right).
\end{align}
$K$ specifies the topological order, while $V$ depends on the microscopic details and sets the edge velocities and their density-density interactions.
We  suppress the indices in what follows.
A generic quasiparticle excitation is characterized by an integer vector $\mathbf{m}$, $\psi_{\mathbf{m}}(y) = e^{i \mathbf{m}^T \phi + i \mathbf{m}^T \mathbf{k} \, y/a}$.
The quantum numbers of the excitation are specified by the `charge vector' $\mathbf{t}$ and `momentum vector' $\mathbf{k}$.
The momentum of $\psi_{\mathbf{m}}$ is $k_{\mathbf{m}} = \mathbf{k}^T \mathbf{m}$, with the convention $k \in [- \pi, \pi]$ regardless of $L$.
The charge of $\psi_{\mathbf{m}}$ is $q_{\mathbf{m}} = \mathbf{t}^T K^{-1} \mathbf{m}$.
As we only probe electronic excitations, we restrict to $q_{\mathbf{m}} = 1$.

	As the bulk is gapped, the dominant long range contributions to $C$ are  power laws from each charge $1$ operator in the edge theory:\cite{wenrev1, moorewen}
\begin{align}
C(x; y) = i \sum_{\mathbf{m}:q_{\mathbf{m}}=1} c_{\mathbf{m}}(x) e^{i k_{\mathbf{m}} y / a} \frac{1}{y^{K(\mathbf{m})}} \frac{1}{|y|^{\eta_{\mathbf{m}}-K(\mathbf{m})}} + \cdots .
\end{align}
Here $K(\mathbf{m})=\mathbf{m}^T K^{-1} \mathbf{m}$ is an odd integer, while $\eta_{\mathbf{m}}$ is the equal time scaling exponent of the excitation $\mathbf{m}$, and depends on both $K$ and $V$.
A consequence of the projection into the LLL is that each $c_{\mathbf{m}}(x)$ is a Gaussian peaked at $x_{\mathbf{m}} = L k_{\mathbf{m}} /2 \pi$; hence the momenta $k_{\mathbf{m}}$ indicates the depth $x$ at which the mode propagates.

	It is easiest to perform the FES collapse in $k$-space.
After Fourier transforming in $y$ to obtain $C(x; k)$, the power-law behavior results in non-analytic dependence on $k$ at the discrete set of momenta $\{ k_{\mathbf{m}} \}$. To proceed we express $C(x; k)$ in the lowest Landau level.
Letting $\psi_n$ denote the field operators of the Landau orbitals, we form the Fourier transformed operators $\psi_k = \frac{1}{\sqrt{N_{\Phi}}} \sum_n e^{-i k n } \psi_n$, where we temporarily consider a finite number of orbitals $N_{\Phi}$.
Note that $\psi_k$ are also the creation operators for orbitals in the $\mathbf{A} = B(0, x)$ gauge convention, localized at $x = L k /2 \pi$.
For large $L/\ell_B$, we find
\begin{align}
\label{Ck}
C(x; k) &= \int dy e^{-i k y / a}  e^{i x y \ell_B^{-2}} \langle \psi^\dagger(x, y) \psi(x, 0) \rangle \\
&= \frac{1}{\sqrt{\pi} \ell_B} n_k e^{- \ell_B^{-2} (L k / 2 \pi - x) ^2} 
\end{align}
where $n_k \equiv \langle \psi^\dagger_k \psi_k \rangle$ is the $k$-space occupation number in the 1D chain.
The gaussian factor implies that the correlations at $x$ are dominated by the behavior near $k \sim 2 \pi x /L$.
In the vicinity of each non-analytic point $n_k$ takes the form \cite{HaldaneRezayi94}
\begin{align}
n_k &\sim  \theta \left( k - k_{\mathbf{m}} \right) |k - k_{\mathbf{m}}|^{\eta_{\mathbf{m}}-1} [ a_0 + a_1 (k - k_{\mathbf{m}}) + \cdots] + \cdots
\end{align}	
where the higher powers of $k$ arise from more irrelevant `descendent' operators.
To determine $\eta_{\mathbf{m}}$ numerically we use a modified version of a fractional derivative defined by
\begin{align}
\mathcal{D}^{\nu} [n_k] = \mathcal{F}[ |r|^\nu \mathcal{F}^{-1}[n_k] ] 
\end{align}
where $\mathcal{F}$ is the Fourier transform.
We expect
\begin{align}
\mathcal{D}^{\eta_{\mathbf{m}}-1}[n_k] &\sim \theta \left( k - k_{\mathbf{m}} \right) [ b_0 + b_1 (k - k_{\mathbf{m}}) + \cdots] + \cdots
\end{align}	
which can be used to check for the correct choice of $\eta_{\mathbf{m}}$.

\begin{figure}[t]
\includegraphics[width=0.5\textwidth]{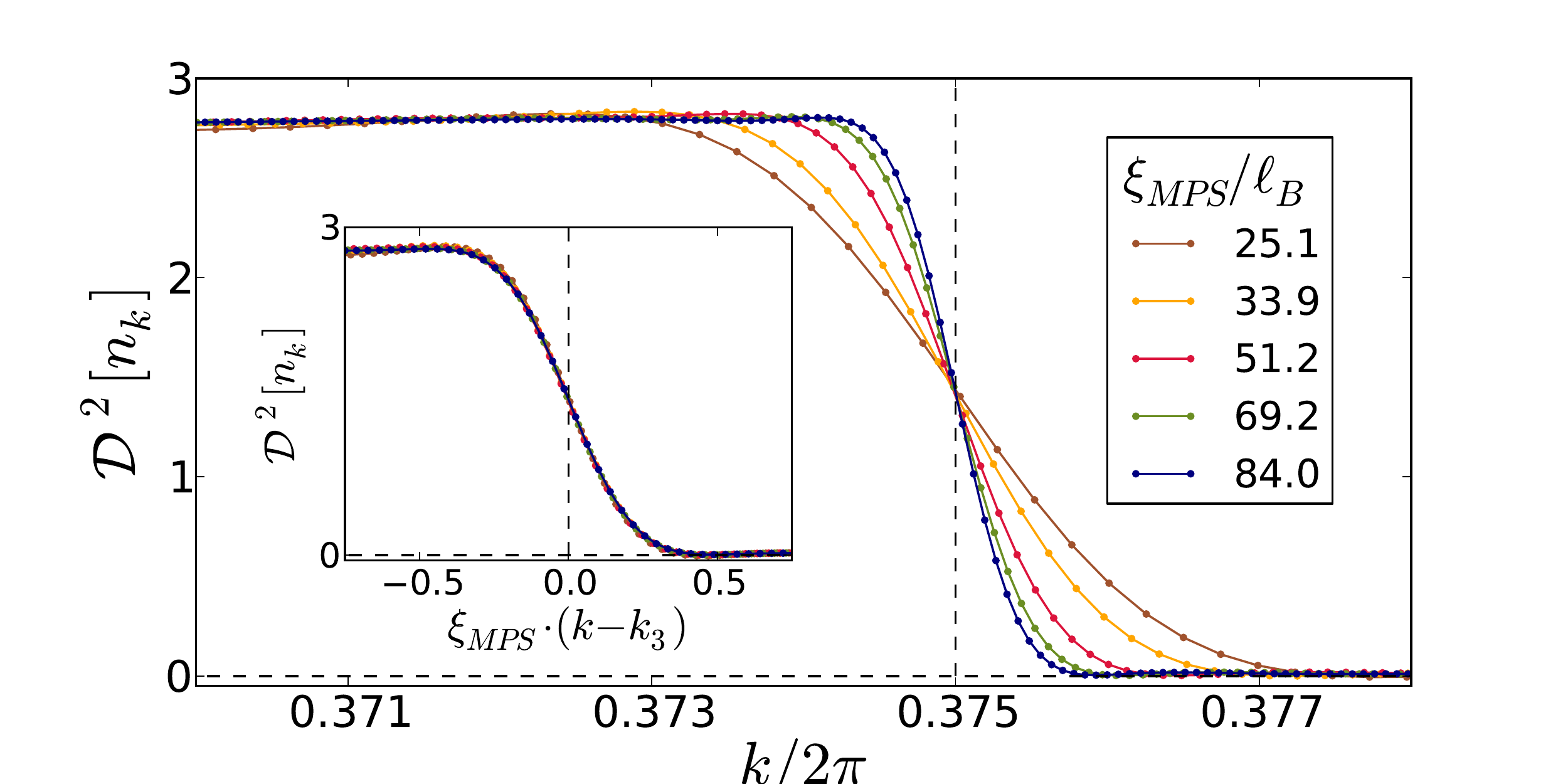}
\caption{Analysis of a $\nu = 1/3$ strip of width $W = \tfrac{3}{4} 17 \ell_B$.
The measured $\mathcal{D}^2[n_k]$ is plotted for increasingly accurate MPSs, parameterized by their correlation length $\xi_{MPS}$.
The crossing of the lines at $k_{3} = \frac{3}{8} \cdot 2 \pi$ indicates a singularity at $k_{3}$, corresponding to the edge of the droplet.
\textbf{Inset}. Scaling collapse supports a true singularity with the predicted exponent $\eta_{3} = 3$.
If $\eta_3 = 3$, then the singular part of $\mathcal{D}^2[n_k]$ is dimensionless near $k_3$, so no vertical scaling is necessary.
The combination $\xi_{MPS}(k - k_3)$ is also dimensionless, so the data should collapse when plotted as a function of $\xi_{MPS}(k - k_3)$.}
\label{fig:q3_collapse} 
\end{figure}

	However, the finite entanglement effects cut off the correlation functions at a scale $\xi_{MPS}$, and hence round out the non-analytic behavior. On dimensional grounds, the smearing must take the form
\begin{align}
\theta(k - k_{\mathbf{m}}) \to s(\xi_{MPS}(k - k_{\mathbf{m}}))
\end{align}
where $s$ is some smoothed version of a step function. 
For the correct choice of $\eta_{\mathbf{m}}$, we can collapse the data by plotting  $\mathcal{D}^{\eta_{\mathbf{m}}-1}[n_k]$ as a function of $\xi_{MPS}(k - k_{\mathbf{m}})$, up to irrelevant corrections and the smooth background, as can be seen in Fig.~\ref{fig:q3_collapse}.
The collapse gives a very precise measurement of both $k_{\mathbf{m}}$ (to better than one part in $10^{-5}$) and $\eta_{\mathbf{m}}$ (to about one part in $10^{-2}$).

\section{Edge universality at $\nu = 1/3, 2/5$ and $2/3$.}
\subsection{The $\nu = 1/3$ edge}
	We first study the filling $\nu = \frac{1}{3}$, a first level hierarchy state whose edge theory supports a single chiral mode.
If the edge is described by a $\chi LL$, the dominant electronic edge exponent is predicted to be quantized to $\eta_3 = 3$ when the two edges of the strip don't interact.\cite{wenrev1}

	We start with a `thick' strip of width $W = \tfrac{3}{4} 17 \ell_B$ on a cylinder of circumference $L = 17 \ell_B$, with a cylinder filling fraction of $\nu_T = 1/4$.
For the interaction we use truncated dipolar repulsion, $V(r) = r^{-3} e^{-(r/7 \ell_B)^2}$, which was used in favor of the Coulomb interaction as a compromise between the increased numerical cost of longer range interactions and the need to ensure the interaction is significantly perturbed from the model Hamiltonian.
Consequently the interaction between the two edges should be very weak.
The distribution $n_k$ is shown in Fig.~\ref{fig:nk}i, including a comparison to the profile when only $V_1$ is used.

\begin{figure}[t]
\includegraphics[width=0.48\textwidth]{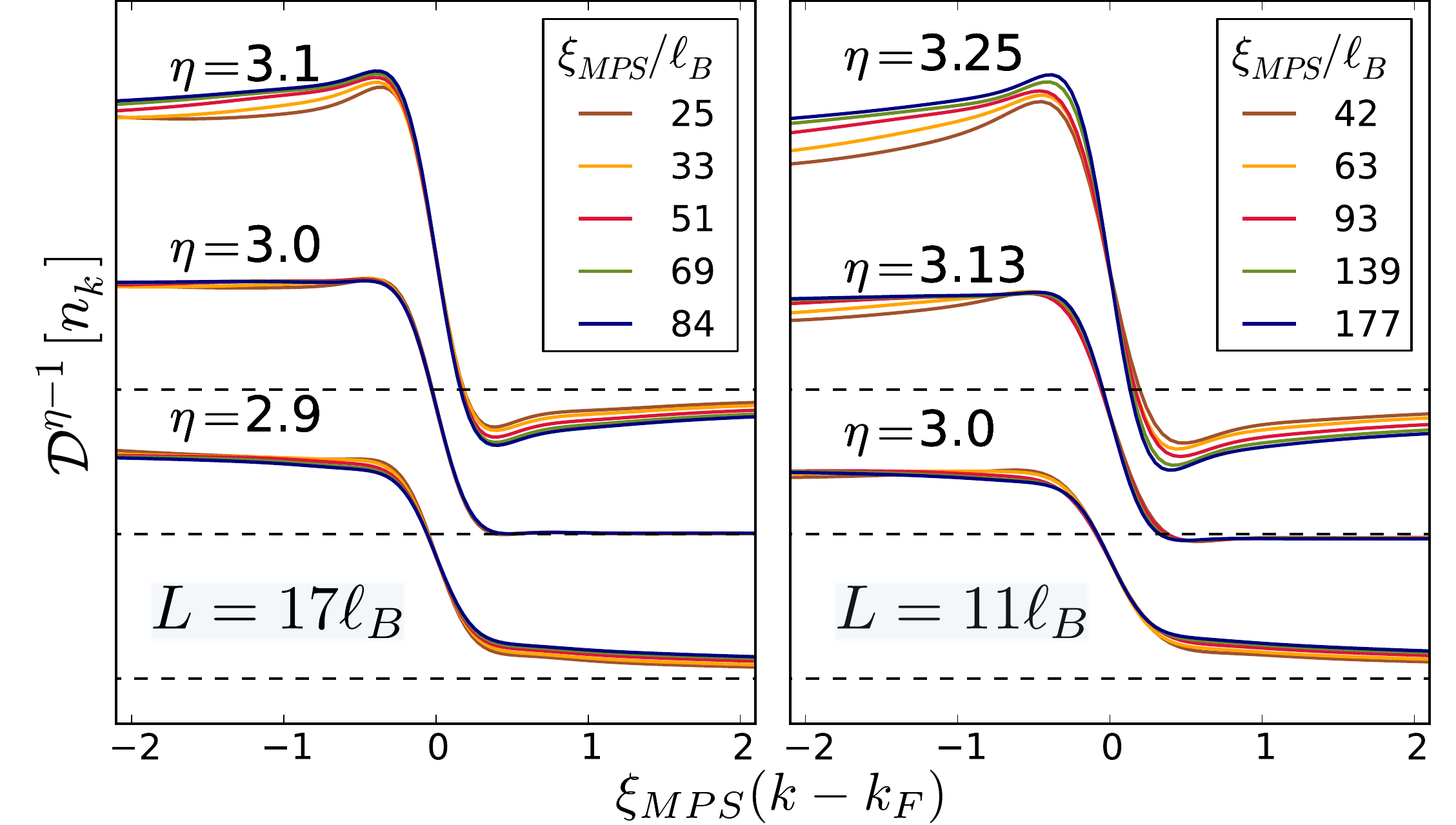}
\caption{Comparing collapse of the $\nu = 1/3$ strip for different ansatz $\eta$. The correct $\eta$ is distinguished by two features: the tightness of the collapse for various $\xi_{MPS}$, and the degree of under/overshoot to the form of a step function. 
The trial collapses at different $\eta$ are shifted apart vertically for clarity.
\textbf{Left panel}) a thick strip, $W = \tfrac{3}{4} 17 \ell_B$. We find $\eta_3 = 3.005 \pm 0.02$, consistent with no inter-edge interactions. \textbf{Right panel}) a thin strip, $W = \tfrac{9}{10} 11 \ell_B$. Because the edges are close to one another, inter-edge interactions renormalize $\eta$ upwards to $\eta = 3.13 \pm 0.02$}
\label{fig:q3_etas} 
\end{figure}

	The dominant singularity of the $\nu = \frac{1}{3}$ edge is observed to occur at $|k_3| = \frac{3}{8}\cdot 2 \pi \pm 10^{-6}$, corresponding to the naive `edge' of the strip as would be obtained from assuming $n_k$ to be a box of height $\frac{1}{3}$, an example of the Luttinger sum rule.
The results of the FES collapse assuming $\eta_3 = 3$ are shown in Fig.~\ref{fig:q3_collapse}, showing excellent agreement.
To check the precision with which we can determine $\eta_3$, we repeat the collapse for various $\eta_3$ as shown in Fig.~\ref{fig:q3_etas}.
The exponent is best fit by $\eta_3 = 3.005 \pm 0.02$.
\begin{figure}[t]
\includegraphics[width=0.48\textwidth]{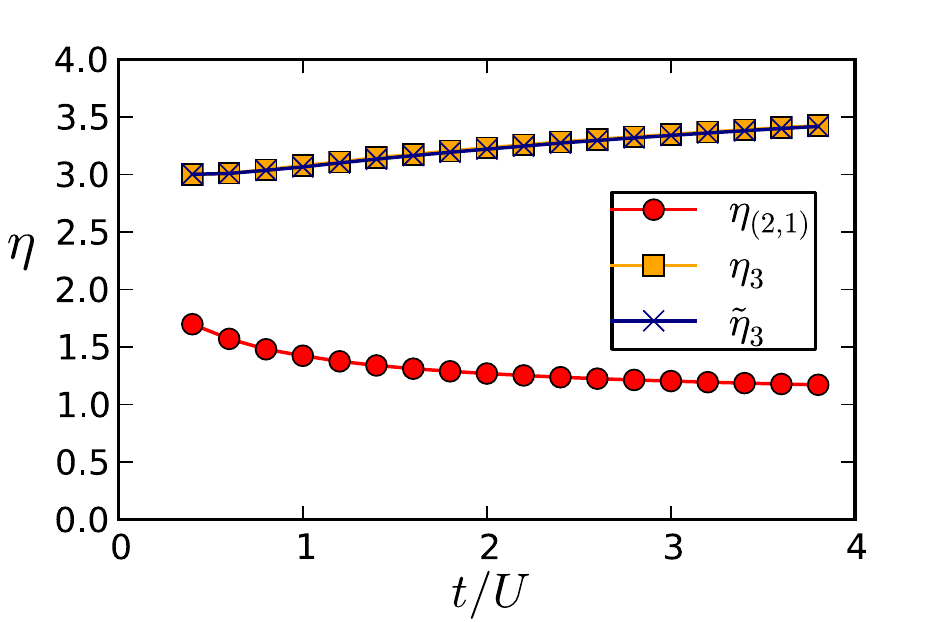}
\caption{Checking the predicted $\chi LL$ exponent relations as the $\nu=1/3$ trap is squeezed.
On a small enough  strip (here $L = 4 \ell_B$) two electronic excitations are visible, corresponding to injecting either $e$ into one edge ( $m=3$) or $2e/3$ into one edge and $e/3$ into the other ($m = (2, 1)$). 
We plot the measured exponents $\eta_3$ and $\eta_{(2,1)}$ as the strength of the trap $t/U$ is increased, as well as the predicted $\tilde{\eta}_3$ given the measured value of $\eta_{(2, 1)}$ using Eq.~\eqref{eq:q3_eta_const}.
At low very $t/U$, $\eta_3 = 3$, the universal $\nu = 1/3$ value.
As the trap is squeezed, the predicted relation of $\chi LL$ theory is satisfied to within 0.5\%.
Note $\eta_{(2, 1)}$ moves towards 1, the exponent of a non-interacting Luttinger liquid.
}
\label{fig:q3_renorm} 
\end{figure}	

\subsection{Renormalization of edge exponents for thin strips}
	To verify that the measurement of $\eta_3 = 3$ is not a bias of the approach, we let the edges interact so that $\eta_3=3$ renormalizes upwards. Using the same interactions and trap profile as before, but using a $L = 11\ell_B, W = \frac{9}{10} 11 \ell_B$ strip, the edges now interact across the vacuum. As shown in Fig.~\ref{fig:q3_etas}, the exponent indeed renormalizes upwards to $\eta_3 = 3.135$.

	For thin strips the $1/3$-state contains multiple electron operators, which correspond to inserting, for example, charge $\tfrac{2}{3} e$ on one edge and $\tfrac{1}{3}e$ on the other, which we label $\mathbf{m} = (2, 1)$.
The amplitude for such a process decays as $e^{-(W/\ell_B)^2/4}$.
$\chi$LL theory predicts a fixed relationship between $\eta_3$ and $\eta_{(2, 1)}$ even if $\eta_3$ has renormalized away from 3, which can be derived in the $K$-matrix formalism.

	A single edge of the $1/3$ state is described by $K = (3)$, $\mathbf{t} = (1)$, but if the two edges of the strip are in proximity we cannot neglect interactions between them.
The full edge theory is $K' = K \oplus (-K)$, $\mathbf{t}' = \mathbf{t} \oplus (-\mathbf{t})$, $\mathbf{k}'=\mathbf{k} \oplus (-\mathbf{k})$ and $V$, restricted by mirror symmetry, has two independent components.
Two singularities we observe correspond to $\mathbf{m} = (3,0)$ and $\mathbf{m} = (2,1)$,\cite{HaldaneRezayi94,SouleThierry2012,Yang2002} and their exponents should satisfy the relation
\begin{align}
\tilde{\eta}_{3} =5 \eta_{(2,1)}-4 \sqrt{\eta_{(2,1)}^2-1}.
\label{eq:q3_eta_const}
\end{align}

	To verify Eq. \eqref{eq:q3_eta_const} we use a thin cylinder $(L = 4 \ell_B)$ so that both excitations are observable.
Keeping the interactions as before, we vary the strength of the trap $t$ relative to the interaction strength $U$.
In Fig. \ref{fig:q3_renorm}, for each $t$ we extract $\eta_3, \eta_{(2, 1)}$ and check their predicted relation.
We find agreement with $\chi LL$ theory to better than 0.5\%.

	In summary, the behavior of the $\nu = 1/3$ edges is well described by $\chi$LL theory, both for thick and thin strips.
For thick strips the edge exponent approaches the quantized value $\eta_3 = 3$.
While we have not simulated the full Coulomb interaction, the interaction is sufficient to significantly perturb the bulk density profile and measurably renormalizes  the exponents when the two edges are close.
It would be interesting to determine whether the quantization $\eta_3 = 3$ is nevertheless a peculiarity either of the $r^{-3}$ interaction or its cutoff at $\sim 7 \ell_B$. 

\begin{figure}[t]
\includegraphics[width=0.5\textwidth]{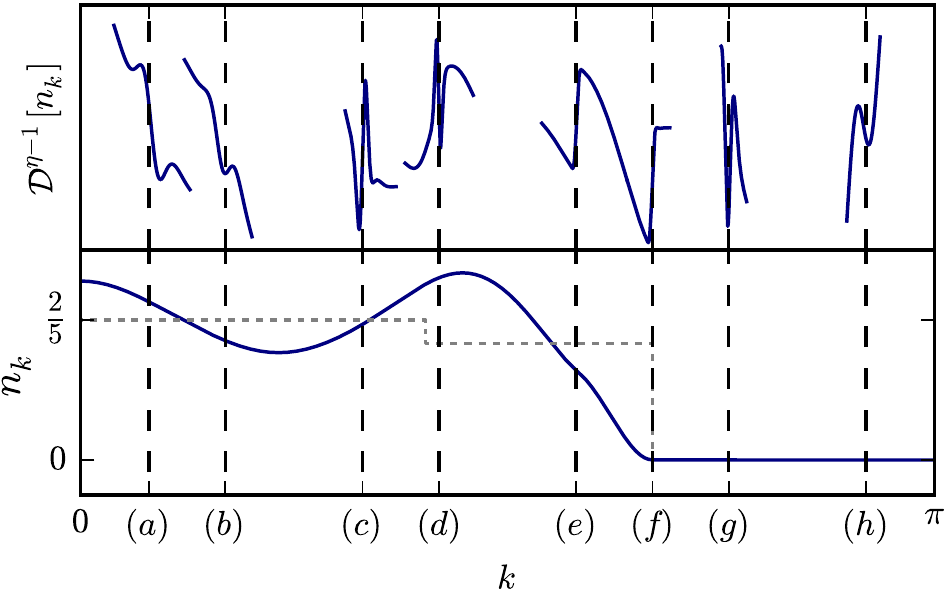}
\caption{The zoo of electron excitations of a $\nu = 2/5$ strip.
\textbf{Bottom panel}. Occupation number $n_k$; vertical dashed lines indicate locations of singularities, only some of which are directly visible in $n_k$.
Dotted line indicates the naive wedding cake density profile.
\textbf{Top panel}. After applying $\mathcal{D}^{\eta -1}$, the singularities appear as step functions.
For the close-up of the $m$th singularity, $\mathcal{D}^{\eta_m -1}$ is applied, some of the regular background is removed, and the vertical axis is rescaled for better visibility.
The two most prominent singularities (e) and (f) have exponents $\eta=3\pm 0.02$, consistent with no inter-edge interactions. The rest of the singularities are found at momenta predicted by the first two. Approximate $\eta$ and $m$ values for the singularities: $\eta_a = 4.8$, $\mathbf{m}_a = (2,-1,2,-2)$; $\eta_b = 5.8$, $\mathbf{m}_b = (1,1,0,2)$; $\eta_c = 5.2$, $\mathbf{m}_c = (1,2,0,1)$; $\eta_d = 4.2$, $\mathbf{m}_d = (3,-2,1,-1)$; $\eta_e = 3$, $\mathbf{m}_e = (2,1,0,0)$; $\eta_f = 3$, $\mathbf{m}_f = (3,-1,0,0)$; $\eta_g = 7$, $\mathbf{m}_g = (4,-3,0,0)$; $\eta_h = 4.2$, $\mathbf{m}_h = (3,0,-1,1)$.}
\label{fig:q25} 
\end{figure}

\subsection{The $\nu = 2/5$ edge}
	The second level hierarchy state at $\nu = 2/5$ has a rich edge structure resulting from the presence of two modes on each edge.
The $2/5$ edge is maximally chiral in the sense that both edge modes propagate in the same direction, so intra-edge interactions are not expected to renormalize the scaling exponents.
In contrast to the $\nu = 1/3$ case, there are multiple ways of inserting charge into a single edge, which appear as a set of singularities.
Using the convention
\begin{align}
K^{-1} &=  \frac{1}{5}\left(\begin{array}{cc}
2 & 1 \\
1 & 3
\end{array}\right) &
\mathbf{t}&= \left(\begin{array}{c}
1 \\
0
\end{array}\right)&
\end{align}
the two most relevant operators are $\mathbf{m}_1 = (2,1)^T$ and $\mathbf{m}_2 = (3,-1)^T$ with $\eta_1 = \eta_2 = 3$.

	We simulate a strip of width $W = \tfrac58 19.2 \ell_B$ on a cylinder of circumference $L = 19.2 \ell_B$, at $\nu_T = 1/4$, using only the hollow-core $V_1$ pseudo-potential.
Adding a small perturbing $V_3$ was not observed to change the exponents.
Our data is consistent with negligible inter-edge interactions, but there is a small amplitude for inserting an electron as a fractional part in both edges.
We can identify a number of small contributions of this type.

	Over two dozen singularities are visible in $n_k$; a summary of the most singular exponents are included in the inset of Fig. (\ref{fig:q25}).
In all cases where the singularity is strong enough to extract $\eta$, it is consistent with the universal values predicted by the $\chi$LL theory.
The dominant exponents $\mathbf{m}_1 = (2,1)$ and $\mathbf{m}_2 = (3, -1)$ are observed to be $\eta_1 = 3.00 \pm 0.02$ and $\eta_2 = 2.995 \pm 0.015$ respectively. 

	The hierarchy picture of the $2/5$ strip is a $1/3$ droplet with an additional condensate of quasielectrons of excess density $1/15$ in the interior.
We cannot directly detect the singularity at the edge of the $1/15$ condensate, 
$\mathbf{m}_3 = (0, 5)$, as the exponent $\eta_3 = 15$ is too large.
Nevertheless, $\mathbf{k}$ can be determined from the momenta $k_1$ and $k_2$ of the two most relevant singularities.
The locations of the remaining $k_{\mathbf{m}}$ are all in agreement with $k_{\mathbf{m}} = \mathbf{k}^T \mathbf{m}$.
Assuming the unobserved $1/15$ edge is at $k_3 = \mathbf{k}^T \mathbf{m}_3$, we find that $\tfrac{1}{3} k_2 + \tfrac{1}{15} k_3 = \nu_T \pi \pm 10^{-5}$.
This is in agreement with the Luttinger sum rule: assuming the $2/5$-state has a `wedding cake'  density profile of $n_k = \tfrac{2}{5}$ for $|k| < k_3$ and $n_k = \tfrac{1}{3}$ for $k_3 < |k| < k_2$, the total electron density is $\frac{1}{3} k_2 + \frac{1}{15} k_3 = \pi \nu_T$.
If the trap potential is modified the $k_{\mathbf{m}}$ change but the constraint is always satisfied.
It is quite remarkable that the naive result is correct to better than one part in $10^{-5}$, as the true density profile has strong oscillations with no actual discontinuities at the $k_{\mathbf{m}}$, as shown in Fig.~\ref{fig:q25}.

\subsection{The $\nu = 2/3$ edge}
	The $\nu = 2/3$ state is also a second level hierarchy state, but the edges are not chiral, and hence the edge exponents are not universal even in the limit of a wide strip.
It has been argued \cite{kfp, kanefisher, wenrev1} that disorder or long-range Coulomb interaction makes the exponents flow to the universal Kane-Fisher-Polchinski (KFP) fixed point, but for our system there is no reason to believe this should happen. 
Using the convention $K = \textbf{diag}(1,-3)$, $\mathbf{t} = (1,-1)$, at intermediate intra-edge interactions the two most relevant excitations of a single edge are $\mathbf{m}_1 = (1, 0)$ and $\mathbf{m}_2 = (2,-3)$.

\begin{figure}[t]
\includegraphics[width=0.48\textwidth]{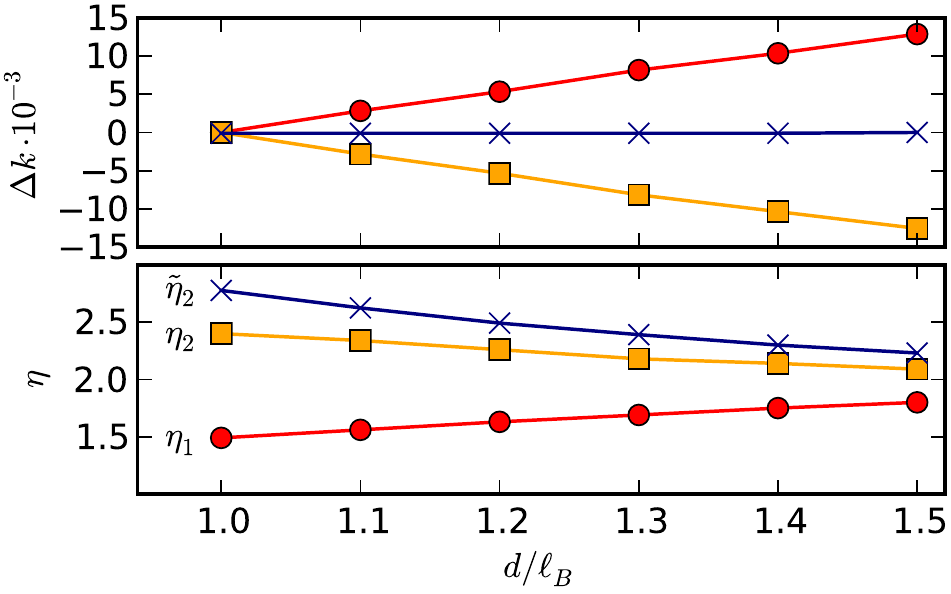}
\caption{(Color online) Renormalization of edge mode momenta and exponents as a function of the edge sharpness $d$ (large $d$ is soft edge).
\textbf{Top panel}: change of momentum $k_1$ (circles) and $k_2$ (squares) with respect to an arbitrary reference. $(k_1 + k_2)/3 - \pi/4$ (crosses) stays $0$ to $10^{-5}$ accuracy, confirming the Luttinger sum rule.
\textbf{Bottom panel}: Exponents $\eta_1$ , $\eta_2$ from iDMRG, and $\tilde{\eta}_2$ calculated from $\eta_1$ assuming $\chi LL$ with no inter-edge interaction.
The significant disagreement between $\eta_2$ and  $\tilde{\eta}_2$ is likely due to inter-edge interactions.}
\label{fig:23_k_eta} 
\end{figure}
	We simulate a $W = 15 \ell_B$ strip on a cylinder of circumference $L = 20 \ell_B$ at $\nu_T = \frac{1}{2}$, using only the hollow-core $V_1$ pseudo-potential. 
In agreement with $\chi LL$ theory, $\eta_1$ is not observed to be quantized.
In order to access different $V$-matrices we can change the sharpness of the edge $d$. 
Under the assumption of vanishing inter-edge interactions, $\eta_1$ and $\eta_2$ should satisfy the $\chi LL$ prediction $\eta_2 = 7 \eta_1-4 \sqrt{3} \sqrt{\eta_1^2-1}$, which they do only to within about $\sim 10\%$, as shown in Fig.~\ref{fig:23_k_eta}.
While the disagreement is likely due to inter-edge interactions, we cannot accurately measure enough of the $\eta_{\mathbf{m}}$ to fully determine the two-edge $V$ to check for consistency.
We again observe a Luttinger sum rule $ (k_1 + k_2)/3 = \pi / 4$.
As the trap is modified the $k_{\mathbf{m}}$ change, but the Luttinger sum rule remains satisfied to one part in $10^{-5}$ accuracy, as shown in Fig.~\ref{fig:23_k_eta}.
It is interesting to note that for softer edges (large $d$), the exponents are observed to renormalize towards the KFP point $\eta_1 = \eta_2 = 2$ (Fig.~\ref{fig:23_k_eta}), even with clean edges and a hollow-core interaction.

\section{Edge states in the 1D picture}
	Our numerics demonstrate robust $\chi LL$ exponents, yet are ultimately realized in a 1D fermion chain.
What can be learned from the 1D point of view?	
First, we apply the Lieb-Schultz-Mattis theorem to the 1D fermion chain to explain the constraint on the $k_{\mathbf{m}}$.
Second, we clarify why strips of different $\nu$, which at first seem to be different phases of matter, can in fact be smoothly deformed to each other.
		
\subsection{A Generalized Luttinger Theorem for Hall Droplets}
	For all three filling fractions the locations of the singularities $k_{\mathbf{m}}$ obey a stringent constraint to better than a part in $10^{-5}$.
In the hierarchy picture, the constraint arises from a caricatured version of $n_k$ which jumps discontinuously to a quantized filling as each level of the hierarchy is added, though $n_k$ looks nothing like this `wedding cake' type profile, as can be seen in Fig.~\ref{fig:nk}.
Letting the jump in filling at the $i$th level of the level-$M$ hierarchy state be $\Delta \nu_i$ at momentum $k_i$, the total filling fraction $\nu_T$ is observed to be 
\begin{align}
\nu_T = 2\cdot \sum^{M}_{i=1} \Delta \nu_i k_i / 2 \pi.
\label{eq:gen_lut_thm}
\end{align}
To our knowledge this `Luttinger sum rule' was first conjectured by Haldane, and taken as the axiomatic starting point of a bosonized description of the excitations. \cite{haldanenotes} 
However, Eq. \eqref{eq:gen_lut_thm} is somewhat unsatisfactory as it appears to single out a particular set $\{\mathbf{m}_i\}$ of the edge singularities out of infinitely many.
There is a natural choice for the hierarchy states, but given a generic $K$ matrix description of an Abelian edge, what constraints are placed on the momenta $k_{\mathbf{m}}$ of the singularities?

	To state the \emph{generalized Luttinger theorem} more precisely, we show that Eq.~\eqref{eq:gen_lut_thm} has the basis independent formulation $\mathbf{k}^T \mathbf{t} = \pi \nu_T$.
We consider only the right edge, as the left has an identical constraint.
The set of singularities $\{\mathbf{m}_i\}$  appearing in Eq.~\eqref{eq:gen_lut_thm} are distinguished as a linearly independent set of $M$ electron operators with trivial mutual statistics,
\begin{align}
\mathbf{m}_i^T K^{-1} \mathbf{m}_j &= \delta_{ij} D_i, &
\mathbf{m}_i^T K^{-1} \mathbf{t} &= 1.
\end{align}
However, there are in fact \emph{multiple} sets $\{\mathbf{m}_i\}$ satisfying this constraint, so we must show that the hypothesis is independent of the choice.
Interpreting $\Delta \nu_i=D_i^{-1}$, the hypothesis reads $\sum_j  D^{-1}_j \mathbf{m}^T_j  \mathbf{k}= \pi \nu_T$. 
Since  $ \mathbf{m}_i^T K^{-1} \left(\sum^M_j  D_j^{-1} \mathbf{m}_j\right) = 1$,
while $\{\mathbf{m}_i^T K^{-1}\}$ is a linearly independent set, we must have $\sum^M_j  D_j^{-1} \mathbf{m}_j = \mathbf{t}$.
Hence the generalized Luttinger Theorem takes the basis independent form $\mathbf{k}^T \mathbf{t} = \pi \nu_T$.

	To prove that $\mathbf{k}^T \mathbf{t} = \pi \nu_T$, we first take a 1D point of view.
Temporarily consider the system on a torus, so that the edges have finite length $L_y$ (in real space). 
According to Ref.~\onlinecite{oshikawaaffleck}, under conditions satisfied by our 1D fermion chain, there exists a low energy ($E\sim 1/L_y$), neutral excitation at crystal momentum $2 \pi \nu_T$.
The non-perturbative proof is an adaption of the Lieb-Schultz-Mattis theorem, ~\cite{lsm} using the `twist operator' $U = e^{2 \pi i \sum_l l \hat{n}_l / N_\Phi}$, where $\hat{n}_l$ is the occupation of orbital $l$ and $N_\Phi$ is the number of orbitals.
As will become clear, we can interpret this excitation as a transfer of charge $\nu$ (the filling fraction of the strip) from the left to right edge, with $\mathbf{m}$ vector $(\mathbf{t}, -\mathbf{t})$.
Accepting this interpretation gives a non-perturbative proof that $(\mathbf{k}^T, -\mathbf{k}^T).(\mathbf{t}, -\mathbf{t}) = 2 \pi \nu_T$, or $\mathbf{k}^T \mathbf{t} = \pi \nu_T$.
Hence Luttinger's theorem for the 1D fermion chain implies the generalized Luttinger Theorem for the Hall fluid.

	To  motivate the identification of the $k = 2 \pi \nu_T$ excitation, we reinterpret this result in terms of the 2D continuum problem. 
The twist operator $U$ acts on the real-space coordinates as translation around the circumference, $(x, y) \to (x + \ell_B^2/L_y, y)$. 
The interaction energy is unchanged, but the trapping energy goes as
\begin{align}
\delta V &= \int dx dy \left[V(x + \ell_B^2/L_y) - V(x) \right] \rho(x, y) \\
&\sim \frac{\ell_B^2}{2 L_y}  \int dx \ell_B^2 V''(x) \rho(x) = \mathcal{O}(1/L_y)
\end{align}
where we have relied on the reflection symmetry $x \to -x$.
Hence the $k = 2 \pi \nu_T$ neutral excitation is simply a small translation of the fluid, which transfers charge from the right to left edge.
To show that the desired excitation is $(\mathbf{t}, -\mathbf{t})$ in the $K$-matrix formalism, recall that threading a $2 \pi$-flux through the cycle $y$ of the torus translates the state by $\ell_B^2/L_y$ in $x$, due to the Hall response of the fluid.
In the bulk, threading $2 \pi$-flux is, by definition, the excitation $\mathbf{m} = \mathbf{t}$.
Since threading flux through $y$ is equivalent to dragging a flux from the left to right edge, the excitation is $(\mathbf{t}, -\mathbf{t})$, as desired.

	For bilayer states, there is a conserved $U(1)$ charge for each component $a$.
A simple extension of the above argument leads to a constraint for each component; if $\mathbf{t}_a, \nu_{T; a}$ is the charge vector and filling of component $a$, then $\mathbf{k}^T \mathbf{t}_a = \pi \nu_{T;a}$.

\subsection{Adiabatic continuity between Abelian edges} 
	In the $\chi LL$ theory, edge theories with different $K$ matrices (modulo an $SL(M, \mathbb{Z})$ equivalence relation) are understood to be distinct phases of matter.
Viewed as a 1D problem, this would seem to imply the existence of distinct classes of metals, even at central charge $c=1$.
We clarify why this is not the case for any finite width strip; in principle all $M$th hierarchy states can be adiabatically continued to one another.
In the 1D picture, this implies they can all be adiabatically transformed to an $M$-component non-interacting metal.
Microscopically, this adiabatic path might require shrinking the cylinder, as for thick strips coupling between the edges is exponentially suppressed.

	The restriction to $SL(M, \mathbb{Z})$ transformations in the $K$-matrix formalism of a single edge is enforced to preserve the compactification lattice of the bosons, which determines the allowed excitations. 
On a geometry with one edge, the edge Hilbert space must contain fractional excitations because we can create a quasiparticle-quasihole pair and bring one particle to the edge while bringing the other infinitely deep into the bulk.
However, if there are two edges, all excitations can be considered edge excitations, and the total topological charge of the two edges together should be trivial.
The restriction to trivial topological charge means a larger class of $SL(2 M,\mathbb{Z}/|K|)$ transformations can be applied, and the distinction between states at the same hierarchy level is lost.

	The Laughlin states, for example,  can all be deformed to the non-interacting IQH state.
In the $\nu = 1/q$ phase,  $K = \mathbf{diag}(q, -q)$, $\mathbf{t} = (1, -1)^T$, $\mathbf{k} = (k, -k)^T$.
By applying an $SL(2,\mathbb{Z}/q)$ transformation
\begin{align}
S = \frac{1}{2}\left(\begin{array}{cc}
q+1 & q-1 \\
q-1 & q+1
\end{array}\right),
\end{align}
we find $\tilde{K}^{-1} = S K^{-1} S^T = \mathbf{diag}(1,-1)$ while $\mathbf{k}$ and $\mathbf{t}$ are unchanged and all the original electronic excitations are spanned by $\mathbf{m} = S \tilde{\mathbf{m}}$ with $\tilde{\mathbf{m}}\in\mathbb{Z}^2$. 
This is the $K$-matrix description of a Luttinger liquid, which implies that any electronic excitation $\mathbf{m}$ at $\nu = 1/q$ can be identified as an excitation of a Luttinger liquid. 
For example, the usual $\mathbf{m} = (3, 0)$ excitation of the $1/3$-state is the $3 k_F$ excitation of a Luttinger liquid, i.e.,  $\mathbf{m} = (3, 0) \leftrightarrow \tilde{\mathbf{m}} = (2, -1)$.
Likewise, the $\mathbf{m} = (2, 1)$ excitation of the $1/3$-state is the usual $k_F$ the of a Luttinger liquid, i.e.,  $\mathbf{m} = (2, 1) \leftrightarrow \tilde{\mathbf{m}} = (1, 0)$. 
By tuning $V$ with the appropriate interactions, we can ensure $\tilde{V} = \mathbf{diag}(\tilde{v}, - \tilde{v})$, so that the exponents will agree with those of non-interacting electrons.

	We have constructed similar explicit transformations for some second and third level hierarchy states.
For certain bilayer states, a similar correspondence is  possible only if we restrict to excitations with integral charge in \emph{each layer separately}, which signifies that such states are only realized with two distinguishable species of fermions satisfying separate charge conservation conditions.

	Adiabatic continuity of this form has already been demonstrated for the thin strip investigated in Fig.~\ref{fig:q3_renorm}. For small $t/U$, $\eta_{3}$ and $\eta_{(2, 1)}$ are close to their quantized $\nu = \frac{1}{3}$ values; as $t/U$ increases and the edges interact, we find $\eta_{(2, 1)} \to 1$, which is the exponent of the $(1, 0)$ excitation of a free Luttinger liquid. 
Throughout the deformation, the functional form of $\eta_3\left( \eta_{(2,1)} \right)$ is as predicted for the $\nu = \frac{1}{3}$ state, and is the \emph{same} as the relation $\tilde{\eta}_{(2, -1)}\left( \tilde{\eta}_{(1,0)} \right)$ of a Luttinger liquid.

\section{Conclusion and future directions}
	In the present work we demonstrated the potential of iDMRG to access the edge physics of FQH phases in a clean infinite strip geometry starting from a microscopic Hamiltonian.
We calculated scaling exponents for multiple edge excitations in the $\nu = 1/3$, $2/5$ and $2/3$ states and found that the predictions of $\chi$LL theory are very accurately met, including the universality of scaling exponents in the maximally chiral $2/5$ edge and their renormalization in non-chiral edges or in the presence of inter-edge interactions. 

	The mapping of the Landau level Hamiltonian onto a fermionic chain offers a 1D point of view on our results.
The occupation number $n_k$ has multiple non-analytic features, which can be identified with edge excitations in the FQH picture.
We demonstrated and analytically proved a long standing conjecture regarding the $k$-values where these features occur, the \emph{generalized Luttinger theorem}, and demonstrated the adiabatic continuity between finite width FQH states and multicomponent Luttinger liquids.

	The techniques used here suggest a number of future directions. 
Of particular interest would be the calculation of exponents in the presence of a point contact, the geometry relevant to interferometry experiments.\cite{Chamon1997}
MPS techniques allow one to introduce a localized defect to the Hamiltonian (a constriction of the trapping potential) while maintaining the infinite boundary conditions of the gapless edge away from the defect. \cite{zaletelmongpollmann, PhienVidalMcCulloch2012}
One could then calculate the inter-edge correlation functions in the presence of an interferometer.  
A second direction would be to investigate more exotic FHQ states, such as the Moore-Read state at filling $\nu = 5/2$, for which significant questions remain regarding the stability of the edge and the interplay between the trap potential and particle-hole symmetry breaking. \cite{WanYangRezayi2006, Wan2008}
Finally, one can apply iDMRG and FES techniques to lattice models on a strip in order to study the edge excitations of other candidate topological phases, either symmetry protected or intrinsic; currently little is known about the microscopics of such edges.

	We thank S.~Parameswaran, R.~Mong, F.D.M.~Haldane, and J.~Jain for useful discussions.
MPZ acknowledges support from NSF GRFP Grant DGE 1106400.
DV and JEM acknowledge support from NSF DMR-1206515.

\bibliography{qh_edges}

\end{document}